\documentclass[aps,prd,nofootinbib,twocolumn,superscriptaddress]{revtex4}

\usepackage{graphicx}
\usepackage{amsmath}

\newcommand{\mrm}[1]{\mbox{\rm #1}}
\newcommand{\be}{\begin{equation}}
\newcommand{\ee}{\end{equation}}
\newcommand{\br}{\begin{eqnarray}}
\newcommand{\bea}{\begin{eqnarray}}
\newcommand{\eea}{\end{eqnarray}}
\newcommand{\er}{\end{eqnarray}}
\newcommand{\ba}{\begin{array}}
\newcommand{\ea}{\end{array}}
\newcommand{\bi}{\begin{itemize}}
\newcommand{\ei}{\end{itemize}}
\newcommand{\bn}{\begin{enumerate}}
\newcommand{\en}{\end{enumerate}}
\newcommand{\bc}{\begin{center}}
\newcommand{\ec}{\end{center}}

\newcommand{\Eq}[1]{Eq.~(\ref{#1})}
\newcommand{\rfn}[1]{(\ref{#1})}

\newcommand{\beq}{\begin{equation}}
\newcommand{\eeq}{\end{equation}}

\newcommand{\U}{\scriptscriptstyle U}
\newcommand{\D}{\scriptscriptstyle D}

\newcommand{\gsim}{\lower.7ex\hbox{$\;\stackrel{\textstyle>}{\sim}\;$}}
\newcommand{\lsim}{\lower.7ex\hbox{$\;\stackrel{\textstyle<}{\sim}\;$}}

\newcommand{\bs}{\begin{small}}
\newcommand{\es}{\end{small}}

\newcommand{\qui}{$q_{{\scriptscriptstyle U}_{\!i}}$}
\newcommand{\qdi}{$q_{{\scriptscriptstyle D}_{\!i}}$}

\def\mysection#1{{\bf #1.} }

\begin{document}

\title{Exponentially spread dynamical Yukawa couplings from non-perturbative chiral symmetry breaking  in the dark sector}

\author{Emidio Gabrielli\footnote{
On leave of absence from Department of Physics, 
University of Trieste, Strada Costiera 11, I-34151 Trieste, Italy.}}
\affiliation{National Institute of Chemical Physics and Biophysics, Ravala 10, 10143 Tallinn, Estonia}
\affiliation{
INFN sezione di Trieste, via Valerio 2, I-34127 Trieste, Italy}

\author{Martti Raidal}
\affiliation{National Institute of Chemical Physics and Biophysics, Ravala 10, 10143 Tallinn, Estonia}
\affiliation{Institute of Physics, University of Tartu, Estonia}

\date{\today}

\begin{abstract}
We propose a new paradigm for generating exponentially spread standard model Yukawa couplings from a new $U(1)_F$ gauge symmetry in the dark sector. 
Chiral symmetry is spontaneously broken among dark fermions that obtain non-vanishing masses from a non-perturbative solution to the mass gap equation.
The necessary ingredient for this mechanism to work is the existence of higher derivative terms in the dark $U(1)_F$ theory, or equivalently the existence of Lee-Wick ghosts,
that $(i)$ allow for a non-perturbative solution to the mass gap equation in the weak coupling regime of the Abelian theory;  $(ii)$ induce exponential dependence of the
generated masses on dark fermion $U(1)_F$ quantum numbers. The generated flavor and chiral symmetry breaking  in the dark sector is transferred 
to the standard model Yukawa couplings at one loop level via Higgs portal type scalar messenger fields. The latter carry quantum numbers of squarks and sleptons.
A new intriguing phenomenology is predicted that could be potentially tested at the LHC, provided the characteristic mass scale of the messenger sector is accessible 
at the LHC as is suggested by naturalness arguments.

\end{abstract}

\maketitle

\section{Introduction}
After the discovery~\cite{Aad:2012tfa} of the Higgs boson~\cite{Englert:1964et} at the LHC,
the only unexplained sector in the standard model (SM) is the flavour sector.
While gauge couplings, such as the electric charge, are fundamental constants of nature following from the
gauge symmetry principle, the SM Yukawa couplings seem not to be connected to any known local or global symmetry. 
Instead, they resemble arbitrary dimensionless numbers spanning over 6 orders of magnitude for charged fermions 
and at least 12 orders of magnitude if the SM neutrinos are Dirac particles. 
All quark flavour and CP-violation experiments over the last 40 years have confirmed the correctness of the SM description of flavour 
observables via the Yukawa interactions~\cite{Buchalla:2008jp}. Lepton flavour observables may indicate 
new physics~\cite{Raidal:2008jk}, such as the seesaw mechanism~\cite{Minkowski:1977sc}, 
but  neutrinos can also be Dirac particles, exactly as the quarks and charged leptons. 
Despite the huge amount of experimental information, constructing  the theory of flavour is one of the biggest challenges
 in modern physics since the physics principles behind it are not known.

 There are only two generic classes of attempts to address the huge spread of the SM Yukawa couplings, each consisting of hundreds of concrete models. 
 The first class is based on the Froggat-Nielsen mechanism~\cite{Froggatt:1978nt}, that introduces $U(1)_F$ flavour symmetric non-renormalizable operators 
  involving a large number of scalar flavon fields that are suppressed by a large cut-off scale $\Lambda.$
  When the flavon $\phi$ obtains a vacuum expectation value (VEV) $\langle\phi\rangle,$
 effective Yukawa couplings $Y$ are induced as powers of the expansion parameter $\lambda\sim \langle\phi\rangle/\Lambda$ 
 as $Y\sim \lambda^n,$ depending on the particle quantum numbers under $U(1)_F$. Since $\lambda\sim 0.2$ to explain the
 Cabibbo angle,  explaining Yukawa couplings within 6 or 12 orders of magnitude requires constructing operators with very high dimensionality.
 It is not clear what underlying physics is responsible for those operators and whether this paradigm is testable.
 
 The second attempt is based on confining different fermions in different branes that are located in different places in extra dimensions~\cite{ArkaniHamed:1999dc}. 
 The Yukawa couplings are induced due to overlaps of the fermion wave-functions with the Higgs wave-function in extra dimensions. 
 This scenario allows for exponential parameterisation of the measured
  Yukawa couplings, but does not explain why their values are what they are. Neither of the attempts is completely satisfactory theoretically,  and none  have any
  experimental support at present.
  
  In this work we propose a new, predictive paradigm for generating exponentially spread SM Yukawa couplings from gauge quantum numbers in the dark sector.
 We assume that in addition to the globally $U(1)^N$ flavour symmetric SM there exists a dark sector with complicated internal dynamics manifested today
 by the existence of dark matter (DM)~\cite{Ade:2013lta}. The origin of flavour symmetry breaking is the chiral symmetry breaking (ChSB)
 due to non-perturbative dynamics in the dark sector. 
 We present  a concrete model with new $U(1)_F$ gauge symmetry in the dark sector that generates masses for dark fermions (singlets under the SM gauge group) 
non-perturbatively {\it a la} the Nambu-Jona-Lasinio (NJL) mechanism~\cite{Nambu:1961tp,Nambu:1961fr}. 
 While the original NJL mechanism operates in the strong coupling regime of the theory,
 our mechanism operates in the weak coupling regime. This is achieved by assuming the existence of Lee-Wick~\cite{Lee:1969fy, Lee:1970iw} type higher derivative terms in the dark
 $U(1)_F$ theory that are equivalent to the existence of negative norm Lee-Wick ghosts \cite{Lee:1969fy, Lee:1970iw,Nakanishi:1971jj, Lee:1971ix,Cutkosky:1969fq}. 
Due to the existence of a massless (or light) dark photon, 
 the generated masses depend exponentially on the $U(1)_F$ quantum numbers~\cite{Gabrielli:2007cp}.
As an additional bonus, when the Lee-Wick theory is generalized to the scalar fields, the Lee-Wick ghosts cancel the quadratic divergences, providing a natural solution to the SM hierarchy problem~\cite{Grinstein:2007mp,Grinstein:2008qq,Grinstein:2008bg, Espinosa:2011js, Espinosa:2007ny}.
 Thus the dynamics and spectacular new features of our mass generating mechanism rely on the Lee-Wick proposal.

 The generated dark fermion  mass spectrum is the source of chiral and flavour symmetry breaking. 
 In our proposal this spectrum is transferred 
to the SM Yukawa couplings at one loop level via Higgs portal type messenger fields, by requiring 
the spontaneous symmetry breaking (SSB) of the discrete Higgs parity symmetry.
Then, the SM Yukawa couplings will be dynamically generated 
in perturbation theory as finite quantities  at one loop.
In addition to dark quantum numbers, the messenger fields must also carry SM quantum numbers that are similar to  the ones of supersymmetric 
 squarks and sleptons. As a result, we obtain effective and finite SM Yukawa couplings of the schematic form
 \bea
 Y^{i}\sim  \exp\left\{-\frac{\gamma}{\alpha q_i^2}\right\},
\label{Yuk}
 \eea
 where $\alpha$ is the  strength of the dark $U(1)_F$ interaction, $q_i$ are the $U(1)_F$ quantum numbers of the fermions $f_i$, $\gamma$ is some universal constant related to the anomalous dimension of the fermion mass operator and 
$i$ denotes flavour. Then, the non-universality of the quark and lepton Yukawa couplings results from the non-universality of the $U(1)_F$ fermion charges $q_i$ in the dark sector.

By means of Eq.(\ref{Yuk}), we are able to explain the exponential spread of SM Yukawa couplings with order one generation dependent $U(1)_F$ charges.
Interesting sum rules are predicted for the mass spectrum as a consequence of Eq.(\ref{Yuk}), which are directly related to the $U(1)_F$ charges.
Incidentally, as we will show numerically, this framework can actually
explain the observed  charged fermions mass hierarchies within a few 
\% level accuracy, by using a simple integer sequence for the  $U(1)_F$ charges.
It can also accommodate the observed patterns of particle mixing.

 The proposed framework predicts rich collider phenomenology that can be potentially tested at the LHC and in future colliders.
 As already stated, the messengers themselves must carry SM quantum numbers similarly to those of squarks and sleptons of supersymmetric theories. 
 If kinematically accessible, those new particles can be produced and discovered at the LHC. Since they couple to the Higgs boson and contribute to the
 Higgs mass at one loop, naturalness arguments require the messenger mass scale to be below 10~TeV. 
 An exact replica of a rescaled SM fermion spectrum is also expected 
in the dark sector, as a consequence of the flavor universality of 
the messenger fields and their couplings to SM
fields.
 The important message to stress is that our scenario is, in principle, directly testable.
 
 We are aware that there is a long way to go towards a more complete 
understanding of the theory of flavour in this approach.  In general, there might be different realisations of both the chiral symmetry breaking 
 mechanism in the dark sector as well as the messenger mechanism presented in this paper. 
 However, we believe that the general features of our proposal are new and could motivate further studies  of dynamical flavour breaking mechanism
  in the presented framework.

The paper is organised as follows. In next section we present details of non-perturbative chiral symmetry breaking in Lee-Wick type models with $U(1)_F$ 
gauge symmetry. The model for generating the SM Yukawa couplings dynamically is presented in section III. In section IV we present the analysis of the
naturalness and vacuum stability bounds.
Phenomenology and direct tests of our
proposal are discussed in section IV. We conclude in section V.

\section{Non-perturbative ChSB mechanism from ${\bf U(1)_F}$ gauge interaction}
In the seminal papers~\cite{Lee:1969fy, Lee:1970iw} Lee and Wick proposed a new
approach to quantum field theories that
prompted the construction of more general theories in which the 
S-matrix is fully unitary, although the Lagrangian is not Hermitian.
This required the introduction of negative norm states which are associated
with massive unstable particles. However, despite the presence of an indefinite 
metric, unitarity can be recovered, provided the
negative norm states are massive and have a finite decay width
\cite{Lee:1969fy, Lee:1970iw,Cutkosky:1969fq}. 
As an advantage, ultraviolet divergences may indeed cancel 
out in the loops due to the indefinite metric of the Hilbert space.
In principle, problems related to  the microscopic violation of Lorentz invariance, that could also arise due to the presence of an 
indefinite metric  in the Hilbert space~\cite{Nakanishi:1971jj},
can be circumvented too~\cite{Lee:1971ix}. 
Indeed, as shown  by Cutkosky {\it et al.}~\cite{Cutkosky:1969fq}, a relativistic and unitary S-matrix can be defined provided 
a new prescription for the deformed energy contour in the Feynman integrals is implemented. Although,
this prescription is not derived from
the first principles of  the  field theory approach, it is well
defined in perturbation theory~\cite{Lee:1970iw,Cutkosky:1969fq}.
There is no rigorous proof yet that the Lee-Wick extensions
could also work at the non-perturbative level. Nevertheless, 
there are studies in this direction leading to a consistent non-perturbative 
approach on the lattice~\cite{Jansen:1993jj,Jansen:1993ji,Fodor:2007fn}.

The original model, satisfying all these requirements,
was the one proposed by Lee and Wick in the framework of 
quantum electrodynamics (QED) \cite{Lee:1970iw}. 
In particular, if one replaces the standard photon field $A_{\mu}$ 
by a complex gauge field $\phi_{\mu}=A_{\mu}+i\, B_{\mu}$,
where $B_{\mu}$ is a massive boson field with negative norm, it is possible
to remove all infinities in QED from the electromagnetic mass differences 
between charged particles. This procedure is equivalent to 
the introduction of
a higher (gauge-invariant) derivative term in the Lagrangian of a
primary $U(1)_F$ gauge field. Then, the mass of the ghost field turns out to be 
proportional to the new physics scale $\Lambda$ 
connected to the higher derivative term.
In order to render charge renormalization finite,
new higher derivative terms should be introduced for the fermion fields
as well.  Although, this procedure shares some similarities with the Pauli-Villars  regularization scheme, in the Lee-Wick approach the massive ghosts 
are not ficticious artefact of the regularization scheme, but
are physical objects associated with observable particle resonances.

Recently, the Lee-Wick approach to a finite theory of QED has been reconsidered in view of  its generalization to the SM. This approach leads to a new SM theory  which is naturally free of quadratic divergencies, thus providing an alternative way to the solution of the hierarchy problem~\cite{Grinstein:2007mp,Grinstein:2008qq,Grinstein:2008bg,Espinosa:2011js}.

\begin{figure}[t]
\begin{center}
\includegraphics[width=0.4\textwidth]{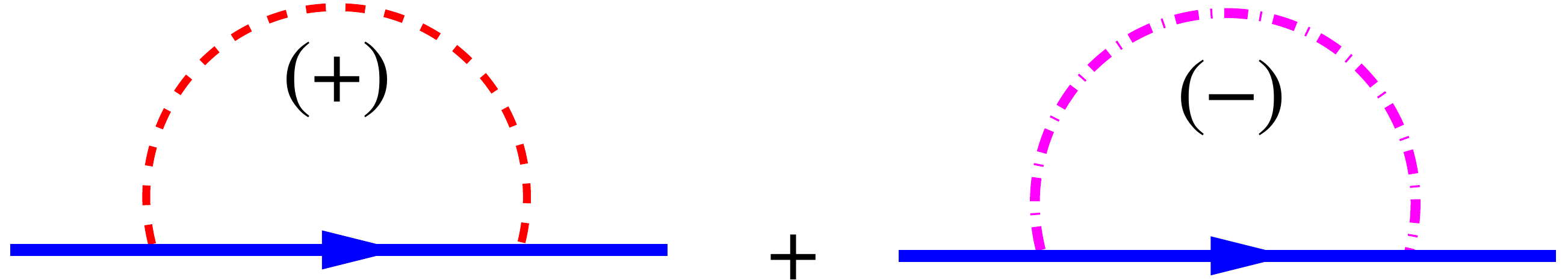}
\caption{One-loop contributions to the fermion self-energy. The dashed and 
dashed-dot lines
indicate the contributions from the positive-norm ($+$) massless and negative-norm ($-$) massive-ghost $U(1)_F$ gauge fields respectively.}
\label{fig1}
\end{center}
\end{figure}

A new interesting feature of the Lee-Wick theories has been recently 
noticed in~\cite{Gabrielli:2007cp}. In particular, if we add 
to a massless Dirac field $\psi$, minimally coupled to a $U(1)_F$ gauge theory,
a higher derivative term in the pure gauge sector of the $U(1)_F$ Lagrangian 
${\cal L}$ as
\bea
{\cal L}& =& -\frac{1}{4}\, F_{\mu\nu}\, F^{\mu\nu}
\,+\, i\bar{\psi}\gamma^{\mu}D_{\mu}\psi\, 
+
\frac{1}{\Lambda^2} 
\partial^{\alpha} F_{\alpha\mu}
\partial^{\beta} F_{\beta}^{~\mu}\, ,
\label{Lag}
\eea
it can be shown that this term can trigger spontaneous chiral symmetry breaking at low energy in the 
weak coupling regime \cite{Gabrielli:2007cp}. In the above equation  $F_{\mu\nu}=\partial_{\mu}A_{\nu}-\partial_{\nu}A_{\mu}$ and 
$D_{\mu}=\partial_{\mu}+i g A_{\mu} $
are the $U(1)_F$ field strength and corresponding covariant
derivative respectively.
This result has been derived by following the approach 
of the NJL mechanism~\cite{Nambu:1961tp,Nambu:1961fr}. In the NJL approach,
the fermion mass term arises as a non-trivial solution of the self-consistent  mass gap equation, namely
\bea
m=\Sigma(\hat{p},m)|_{\hat{p}=m}\,   ,
\label{gap-eq}
\eea
where $\Sigma$ stands for the fermion self-energy induced by the interaction.
Now, due to the presence of the indefinite metric and $U(1)_F$ gauge symmetry
for the Lagrangian in Eq.(\ref{Lag}), the self-energy $\Sigma$ turns out to be 
finite at one loop and gauge invariant~\cite{Gabrielli:2007cp}. By computing 
the Feynman diagrams in Fig.\ref{fig1} for the self-energy, 
and implementing 
the mass-gap equation in Eq.(\ref{gap-eq}), we obtain
\bea
\!\!\!\!\!m&\!=\!& -\frac{\alpha m}{2\pi}\! \int_0^1 \!\! dx (2-x)
\log{\left(\frac{m^2}{\Lambda^2}\frac{(1-x)^2}{x}\right)} + 
{\cal O }( \frac{m^2}{\Lambda^2}) \, .
\label{gap2}
\eea
In the above equation  we neglected terms of order ${\cal O }( m^2/\Lambda^2)$ 
since we are interested to see if there is a non-trivial 
mass-gap solution corresponding to the  case in which $m\ll \Lambda$.
As we can see, this equation admits two solutions. One trivial, corresponding to $m=0$ and related to the perturbative vacuum, and a non-trivial one $m\neq 0$
corresponding to the non-perturbative vacuum. Following the arguments 
exposed in~\cite{Nambu:1961tp,Nambu:1961fr}, it can be shown that 
the vacuum state associated with the minimum energy, 
is the one corresponding to the massive solution. Hence, the true vacuum
corresponds to the phase of ChSB and  is 
orthogonal to the perturbative vacuum.

Finally, by solving the mass-gap equation in (\ref{gap2}) at the leading order, we get the following the results \cite{Gabrielli:2007cp},
\bea
m=\Lambda \exp\left\{-\frac{2\pi}{3\alpha} +\frac{1}{4}\right\}\, ,
\label{mgap}
\eea
where $\Lambda$ is the scale associated with the higher derivative term and 
$\alpha=g^2/4\pi $ is the effective 
fine-structure constant. In order to include
the resummation of the leading log terms  $\alpha^n \log^n(\Lambda/m)$,
expected to come from higher order contributions in perturbation theory, 
$\alpha$ appearing in Eq.(\ref{mgap}) should be substituted 
with the running coupling constant 
$\alpha(\Lambda)$ evaluated at the high energy scale $\Lambda$, namely 
\footnote{Notice that, in \cite{Gabrielli:2007cp} $\alpha$ has been set at the
scale $m$ in the corresponding solution for the mass-gap equation, 
missing the proper resummation of the leading log terms. This led to an inconsistent condition, namely that this solution was allowed only  for $N_f<2$, with $N_f$ the number of fermions charged under $U(1)_F$, which was just a consequence of the incorrect scale at which $\alpha$ inside Eq.(\ref{mgap}) was evaluated.}
\bea
m=\Lambda \exp\left\{-\frac{2\pi}{3\alpha(\Lambda)} +\frac{1}{4}\right\}\, .
\label{mgap-resum}
\eea
This relation can be also be expressed  as a function of $\alpha(\mu)$ evaluated at an arbitrary  renormalization scale $\mu < \Lambda$, as follows
\bea
m&=&\Lambda \exp{\left\{-\frac{2\pi}{3\alpha(\mu)}+\frac{1}{4}\right\}}
\left(\frac{\Lambda}{\mu}\right)^{\frac{4}{9}}
\, ,
\label{msol2}
\eea
where the $U(1)$ one-loop beta-function has been used. It is easy to check that the r.h.s. of Eq.(\ref{msol2}) is independent on $\mu$, consistently with the beta-function evaluated at the leading order in $\alpha$.
As we can see from the exponential 
dependence of the coupling constant $\alpha$, 
the solution in Eq.(\ref{mgap-resum}) is a truly non-perturbative one.
However, notice that this solution always exists in the weak coupling regime, 
$\alpha \ll 1$, since its consistency requires that $\alpha \ll 8\pi/3$.
Remarkably, in the original NJL solution, derived
by introducing {\it ad hoc}  chiral symmetric four-fermion contact
interaction, a strongly coupled regime was required to break the chiral symmetry.
A generalization of this result for the corresponding 
non-abelian $SU(N)$ interaction can be found in~\cite{Gabrielli:2007cp}.

The main difference between the solution in (\ref{mgap})-(\ref{mgap-resum}) and the corresponding
one in the NJL model is due to the fact that in our case the fundamental interaction has a  $U(1)_F$ local gauge symmetry. The fact that there exists a non-trivial 
mass solution is actually peculiar to the $U(1)_F$ and $SU(N)$ gauge interactions and does not hold in general. Indeed, in the case in which  the chiral invariant interaction is replaced by a massless scalar and pseudoscalar 
fields coupled to the fermion field in a chiral invariant way, 
supplied by a higher derivative term in the kinetic part of the scalar Lagrangian, 
the non-trivial mass solution does not exists in the weak coupling
regime~\cite{Gabrielli:2007cp}.
Indeed, in this case, the corresponding sign in front of the integral in Eq.(\ref{gap2}) turns out to be positive. This suggests that the existence of the
non-trivial solution in Eq.(\ref{mgap}) is related to the spin-1 nature 
of the field that generates the long distance interaction in the fermion 
sector.

If the presence of the Lee-Wick term
is a common feature of unbroken gauge theories, then there should also be 
a contribution to the SM fermion masses 
induced by the Lee-Wick term of QED. However, this effect is totally 
negligible in QED, assuming the corresponding scale 
$\Lambda$ lower than the Planck scale $M_{\rm Pl}$.
Indeed, for
$\Lambda \sim M_{\rm Pl}$, the corresponding mass contribution to the charged leptons is of order of $10^{-97}$ eV.
For the analogous effect in QCD, see \cite{Gabrielli:2007cp}.

The generalization of  Eqs.(\ref{mgap})-(\ref{mgap-resum}) to $N_f$ 
fermions coupled to 
a $U(1)_F$ gauge field with different $q_f$ charges is straightforward. 
Let us consider now the same Lagrangian as in Eq.(\ref{Lag}), but with 
the fermionic term replaced by
\bea
{\cal L}_F=i\sum_{f=1}^{N_f}\bar{\psi}_f\gamma^{\mu}\left(\partial_{\mu}+i g 
\hat{Q} A_{\mu}
\right) \psi_f\, ,
\label{Lag2}
\eea
where $\hat{Q}$ is the $U(1)_F$ quantum charge operator satisfying the 
relation $\hat{Q}\psi_f=q_f\psi_f$. The total Lagrangian is now invariant under the generalized $U(1)_F$ gauge transformations
\bea
\psi_f ~\to~ e^{i\varepsilon(x)\, q_f} \psi_f\, ,\, ~~~
A_{\mu} ~\to~ -\frac{1}{g}\partial_{\mu} \varepsilon(x),
\label{gauge}
\eea
where $\varepsilon(x)$ is the usual local gauge parameter.
Then, the result in Eq.(\ref{mgap-resum}) is generalized to
\bea
m_f=\Lambda \exp\left\{-\frac{2\pi}{3\alpha(\Lambda) q_f^2} +\frac{1}{4}\right\}\, .
\label{mgap2}
\eea
As we can see, the mass degeneracy in the $N_f$ fermion system 
is now removed by the splitting among the $U(1)_F$ charges. 
We can get an exponential mass spread, with the exponential argument being
proportional to the inverse square of the quantum charges. 
The small breaking of the global $SU(N_f)$ symmetry is exponentially amplified by the mass spectrum 
generated by the non-perturbative ChSB mechanism. 
However, notice that the Lagrangian Eq.(\ref{Lag2}) is still $U(1)_F$ gauge invariant after 
the spontaneous ChSB, since the fermion mass matrix is a function of the charge operator 
$\hat{Q}$.

Notice that the solution in Eq.(\ref{mgap-resum}) implies
a relation between $\alpha(\Lambda)$ and $\alpha(m)$, which in the case of $N_f$ fermions with unity charge minimally coupled to $U(1)_F$, is given by
\bea
\alpha(\Lambda)=\alpha(m)\left(1+\frac{4}{9}N_f\right) \, ,
\eea
where in deriving the above expression the  $U(1)$ $\beta$-function 
at one-loop has been used and  the $1/4$ factor inside the exponent of
Eq.(\ref{mgap-resum}) has been neglected in the weak coupling regime
$\alpha(\Lambda)\ll 1$.

Now, it is tempting to speculate whether this ChSB pattern for the fermion masses
could be consistent with the observed mass spectrum of quarks and leptons.
Let us consider first the charged lepton mass spectrum.  Due to the mass hierarchy in 
Eq.(\ref{mgap2}), we should expect $q_{e} > q_{\mu} > q_{\tau} $. 
For example, we can extract the values of $\alpha$ and $\Lambda$ from the measured masses
and assumed $U(1)_F$ charges of the electron and muon, namely
\bea
\alpha^{-1} &=&\frac{3}{2\pi}\frac{q_e^2q_{\mu}^2 
\log{\left(\frac{m_e}{m_{\mu}}\right)}}{q_e^2-q_{\mu}^2},
\nonumber \\
\Lambda &=& m_{e}\left(\frac{m_{\mu}}{m_{e}}\right)^{
\frac{q_{\mu}^2}{q_{\mu}^2-q_{e}^2}}\, .
\eea
If we assign, for example, the charges in the lepton sector 
as a sequence of integer numbers as
 $q_{e}=4,~q_{\mu}=5,~q_{\tau}=6$, 
we get
\bea
\alpha^{-1}(\Lambda) &\simeq& 113, ~~ \mrm{and} ~~
\Lambda\simeq 1.4 ~{\rm TeV}.
\label{leptoncoupling}
\eea
This will give the following prediction for the tau lepton mass
\bea
m_{\tau} &\simeq& 1.9~{\rm GeV}.
\eea
Thus, this charge pattern gives the mass of the $\tau$ lepton 
within 7\% accuracy, without invoking any order one coefficient that, in principle, could be present. 

Similarly, assuming Dirac neutrino masses with $q_{\nu_\tau}=3$ as the charge of the heaviest light neutrino, 
and using exactly the same values for the interaction strength and the new physics scale as for the charged leptons
given by \Eq{leptoncoupling}, we obtain a prediction for the neutrino mass scale
\bea
m_{\nu_{\tau}}&\simeq & 5~{\rm eV}\, .
\eea
Although this value is just a bit too large to be consistent with the direct neutrino mass measurements and with cosmological constraints on 
the neutrino mass scale, it might not to be totally unrealistic. 
However, as we will show in the following, the tree-level 
coupling of $U(1)_F$ to SM fermions cannot be a realistic model, and 
this problem would require a different implementation of the main idea.

Incidentally, for the quark spectrum, we found that a good fit is obtained
also following the sequence of 4, 5, 6, integer charges separately for up- and down-quark sectors.
The corresponding mass predictions are within 20-40\% accuracy. 
This indicates that additional corrections of order ${\cal O}(1)$ are needed in the quark sector, as is the case also for the 
Frogatt-Nielsen mechanism. 

Clearly, this example should not be taken as a realistic model of flavor,
since there are several flaws showing that it cannot be
phenomenological acceptable. First of all, an exact $U(1)_F$ gauge interaction
cannot be simply coupled at tree-level with SM fermions, unless it is
extremely weak, which is not the case here. Notice that in the above example
the effective strength at the $\Lambda$ scale 
of this new interaction coupled to electrons is
$q_e^2\alpha(\Lambda)\sim 0.14$, with $q_e=4$, which is 
almost twenty times stronger than EM interactions.
Second, if we require that quarks and lepton masses arise
from the SM Higgs mechanism, as is confirmed by the global fits to the LHC and Tevatron data~\cite{Giardino:2013bma}, 
this extra contribution to their masses 
would spoil the tree-level relation of SM Yukawa couplings with masses,
and eventually the unitarity of the SM.

However, we will see that there is actually a phenomenologically
viable way to implement this mechanism to generate the hierarchy of 
the SM fermion masses.
The main idea is to assume that this mechanism is acting on fundamental 
fermions which belong to a dark sector. These fermions must be singlet under the SM gauge group.
Then, the flavor and ChSB of the dark sector is transferred to the SM Yukawa couplings by Higgs portal type messenger fields. 
We will see that the Yukawa couplings can be actually generated by finite radiative corrections and  be proportional to the masses of dark fermions. The
latter ones play now the role of a primary source of flavor and ChSB in the
SM. In the next section we shall present a model for the messenger sector,
and provide predictions for the finite one loop SM Yukawa couplings.

\section{Generation of the SM Yukawa couplings from dark dynamics}

In this section we present the
Lagrangians for the dark and messenger sectors, following
the model building guidelines of the previous section, and compute the induced SM Yukawa couplings. 
Let us start with the dark sector.

The dark sector is assumed to be composed by Dirac fermions  $Q^{U_i,D_i}$ 
which are similar to replica of the SM fermions, although they are singlet 
under the SM gauge interactions, where $i,j$ indicate the flavor. 
We will focus here only on the quark sector, the generalization 
to the leptonic sector will be straightforward.
These fermions are assumed to be massless at tree-level and satisfy an
exact dark $U(1)_F$ gauge symmetry. The pure gauge sector will be supplemented  by a 
Lee-Wick term as in Eq.(\ref{Lag}), in order to dynamically trigger spontaneous ChSB at low energy.
Notice that chiral symmetry is assumed to be an exact
symmetry of the Lagrangian, that is spontaneously broken
by the $U(1)_F$ gauge interaction. This assumption avoids to introduce generic tree-level mass terms for the fermions, that would explicitly break chiral symmetry and eventually spoil the predictions of exponentially spread mass gaps.

Then, the Lagrangian of the dark sector is given by
\bea
{\cal L}_{DS}&=& 
i\sum_i \left( 
\bar{Q}^{U_i}{\cal D}_{\mu}\gamma^{\mu} Q^{U_i}+\bar{Q}^{D_i}{\cal D}_{\mu}\gamma^{\mu} Q^{D_i}\right)
\nonumber \\
&+&
\frac{1}{4} F_{\mu\nu} F^{\mu\nu} - \frac{1}{\Lambda^2}  \partial^{\mu} F_{\mu\alpha} \partial_{\nu} F^{\nu\alpha},
\label{LagDS}
\eea
where ${\cal D}_{\mu}=\partial_{\mu}+i g \hat{Q} A_{\mu}$
is the covariant derivative associated with 
the $U(1)_F$ gauge field, with $\hat{Q}$ the charge operator acting on the fermion 
fields $Q^{U_i}$, $Q^{D_i}$, and $F_{\mu\alpha}$ is the corresponding $U(1)_F$ field strength tensor. In order to explain the large mass splitting, 
we assume that the $U(1)_F$ quantum charges are not degenerate, 
and indicate them with \qui, \qdi 
corresponding to the fields  $Q^{U_i}$, $Q^{D_i}$ 
respectively. The Lagrangian Eq.(\ref{LagDS}) is invariant under the 
corresponding $U(1)_F$  gauge transformations given in Eq.(\ref{gauge}).
Therefore, the flavor symmetry is explicitly broken by 
the non-universality of $U(1)_F$ quantum charges.

Given the particle content and the $U(1)_F$ gauge interaction in the dark sector, the dark fermions obtain masses as described in the previous section.
Those generation dependent masses are exponentially spread according to their gauge quantum numbers. We assume that this is the origin of 
chiral and flavour symmetry breaking in nature that is communicated to the SM.
As we will see,  this will necessarily require us to introduce Higgs portal type interactions, mediated by scalar messenger fields. 

Basically, the main idea is the following. We assume that the 
SM structure remains the same 
at low energies, while the Yukawa couplings should emerge (as finite contributions) at one loop order due to the interaction of the SM fields with the messenger
sector. The SM fermions acquire mass by means of the SM Higgs
mechanism, but now the splitting between the Yukawa couplings 
is naturally explained by the large mass differences of fermions 
in the dark sector. Although there might be other ways to implement the messenger sector 
for generating finite Yukawa couplings, the requirements of both having finite Yukawa couplings at one loop and renormalizable 
$SU(2)_L\times U(1)_Y\times U(1)_F$ 
invariant interactions in the messenger sector, will strongly reduce many other 
potential choices.

Before entering into the details of the structure of the messenger sector, 
we would like to discuss some relevant issues.
A crucial constraint that must be  imposed in order to 
radiatively generate the Yukawa 
couplings is to avoid the presence of SM Yukawa couplings at the tree-level. 
This can by simply achieved by imposing a discrete Higgs parity symmetry, namely $H\to -H$, where $H$ stands for the SM Higgs boson doublet under 
$SU(2)_L$. Indeed, the SM Yukawa couplings are the only interaction terms in the SM in which $H$ appears linearly. Therefore forbidding the SM Yukawa interaction
terms is technically natural.

In order to generate (finite) Yukawa couplings at one loop level, this parity symmetry must be broken. Thus we need to introduce a singlet scalar field 
$S_0$ (under SM gauge 
interactions) which is coupled to the Higgs field, and that transforms as 
$S_0\to -S_0$ under the Higgs parity transformation $H\to -H$.  This implies
that the Yukawa couplings are always proportional to the VEV $\mu=<S_0>$ of the 
scalar field associated with the SSB of this discrete symmetry
\footnote{The spontaneous breaking of discrete $Z_2$ symmetry may generate
cosmological problems because of domain walls. A solution is 
to break the discrete symmetry explicitly with a small parameter so that
the model features are not changed numerically~\cite{Sikivie:1982qv}. Here we assume that
the $Z_2$ symmetry is explicitly broken by small parameters in the scalar
sector of the model, like $\rho  S H^2$, with $\rho\ll M_H,$ that does not change our results. 
Alternatively, if the scale of inflation is below $<S_0>$, the domain walls are diluted by inflation.
In the following we assume that the domain wall problem is solved in our model. }.

In the case in which the scalar messenger  masses are much larger than the 
dark fermion masses, by using dimensional analysis, the generated Yukawa couplings are expected to be of the form
\bea
Y_i \sim \frac{M_{Q_i} \mu L}{\bar{m}^2}\, ,
\label{Y}
\eea
where $M_{Q_i}$ is the mass of the dark fermion, which plays the role of the
primary ChSB source, and $\bar{m}$ is an average mass of the messenger fields.
Here $L$ is a dimensionless constant, expected to be $L \ll 1$, which absorbs
all loop factors and products of perturbative coupling constants in the
messenger sector.
While $M_{Q_i}$ and $\bar{m}$ are masses of dynamical particles, 
the singlet VEV  $\mu$ is an external mass scale. The latter property allows us to have an
extra free parameter necessary 
for adjusting the normalization of $Y_i$ at the right 
phenomenological scale.
Then, we can see that the hierarchy of fermion masses $M_{Q_i}$ in the dark
sector is directly translated to 
the hierarchy of SM Yukawa couplings, provided the scalar messenger sector is
heavier than the dark fermion one. A similar  conclusion is achieved in the opposite case when $M_{Q_i} \gg \bar{m}$. In this case the 
scaling properties of Eq.(\ref{Y}) should be replaced by 
\bea Y_i \sim \frac{\mu L}{M_{Q_i}}\, ,
\eea
reversing the hierarchy of the Yukawa couplings as a function of the dark 
fermion masses. As we will show in the following, 
the latter realization would be phenomenologically disfavoured since, due
to the conservation of the $U(1)_F$ charge, some messenger fields that are 
charged under the SM gauge group might become stable.

The total tree-level Lagrangian can be expressed as follows
\bea
{\cal L}&= {\cal L}^{ Y=0}_{SM} + {\cal L}_{MS} + {\cal L}_{DS},
\eea
where  ${\cal L}^{Y=0}_{SM}$ is the SM Lagrangian with vanishing
tree-level Higgs Yukawa couplings, ${\cal L}_{MS}$ is the Lagrangian 
containing the messenger sector with its couplings to the SM and dark fields,
and ${\cal L}_{DS}$ is the Lagrangian in Eq.(\ref{LagDS}). 
The ${\cal L}_{MS}$ Lagrangian communicates the ChSB of the dark sector to the SM observable one through the generation of Higgs Yukawa couplings at one loop.

In order to have a Higgs portal type messenger sector, which is invariant under the SM gauge group and under the $U(1)_F$ gauge theory, the minimum
set of messenger fields required is the following
\begin{itemize}
\item $2N_f$ complex scalar $SU(2)_L$ doublets: $\hat{S}_L^{\U_i}$ and $\hat{S}_L^{\D_i}$,
\item $2N_f$ complex scalar  $SU(2)_L$ singlets: $S_R^{\U_i}$ and $S_R^{\D_i}$,
\item one real $SU(2)_L\times U(1)_Y$ singlet scalar: $S_0$,
\end{itemize} 
where
$\hat{S}_L^{\U_i,\D_i}=\left(\begin{array}{c}S^{\U_i,\D_i}_{L_1}\\S^{\U_i,\D_i}_{L_2}
\end{array}\right)$, 
$N_f=3$ and $i=1,2,3$ stand for the flavor index.
It is understood that the messenger and corresponding dark fermion fields 
associated with the leptonic sector will follow the same pattern as for the
quark sector, assuming that neutrinos are of Dirac type.
In the following we will discuss only  the quark sector,  the
extension to the leptonic sector will be straightforward.

Notice that the messenger fields $\hat{S}_{L}^{\U_i,\D_i}$, $S_{R}^{\U_i,\D_i}$ carry the SM quantum numbers of quarks, where the labels $L,R$ stand for 
the corresponding  chirality structure of the SM fermions. 
Therefore, they couple both to  the electroweak gauge bosons and to the gluons in the standard way.  
In this respect they resemble the squarks of the  supersymmetric extensions of the SM. Analogous conclusions  hold in the case of 
extensions of the messenger field content to the lepton sector.

The quantum numbers of the messenger fields are reported in Table~\ref{tab1}. 
Corresponding entries in the columns of $SU(2)_L$ and $SU(3)_c$ 
refer to the group representations, namely $1/2$ and $3$ 
for doublets and triplets respectively, while the entries in 
$U(1)_{Y}$ and $U(1)_{F}$ columns stand for the corresponding
quantum numbers for hypercharge $Y$ and $U(1)_F$ dark sector $q_f$,
respectively. The electromagnetic (EM) quantum charges, in units of 
the electric charge $e$, are given by 
the SM relation $Q_{\rm EM}=t_3+Y/2$, with $t_3$ the corresponding 
eigenvalue of the $SU(2)_L$ diagonal generator, namely 
$t_3(S^{\U_{i},\D_{i}}_{L_1})=1/2$, $t_3(S^{\U_{i},\D_{i}}_{L_2})=-1/2$, and 
$t_3(S^{\U_{i},\D_{i}}_{R})=0$.
\begin{table} \begin{center}    
\begin{tabular}{|c||c|c|c|c|c|}
\hline 
Fields 
& Spin
& $SU(2)_L$ 
& $U(1)_Y$
& $SU(3)_c$
& $U(1)_F$
\\ \hline 
$\hat{S}_L^{\D_i}$
& 0
& 1/2
& 1/3
& 3
& -\qdi
\\ \hline
$\hat{S}_L^{\U_i}$
& 0
& 1/2
& 1/3
& 3
& -\qui
\\ \hline
$S_R^{\D_i}$
& 0
& 0
& -2/3
& 3
& -\qdi
\\ \hline
$S_R^{\U_i}$
& 0
& 0
& 4/3
& 3
& -\qui
\\ \hline
$Q^{\D_i}$
& 1/2
& 0
& 0
& 0
& \qdi
\\ \hline
$Q^{\U_i}$
& 1/2
& 0
& 0
& 0
& \qui
\\ \hline
$S_0$
& 0
& 0
& 0
& 0
& 0
\\ \hline \end{tabular} 
\caption[]{
Spin and gauge quantum numbers for the messenger fields. The group 
$U(1)_F$ corresponds to the gauge symmetry group of the dark sector.
}
\label{tab1}
\end{center} \end{table}

Finally, for the interaction Lagrangian   ${\cal L}^I_{MS}$ 
of the messenger sector with quarks and SM Higgs boson we have
\bea
{\cal L}^I_{MS} &=&
g_L\left( \sum_{i=1}^{N_f}\left[\bar{q}^i_L Q_R^{\U_i}\right] \hat{S}^{\U_i}_{L} +
\sum_{i=1}^{N_f}\left[\bar{q}^i_L Q_R^{\D_i}\right] \hat{S}^{D_i}_{L}\right)+
\nonumber\\
&+&
g_R\left(\sum_{i=1}^{N_f}\left[\bar{\scriptstyle U}^i_R Q_L^{\U_i}\right] S^{\U_i}_{R} +
\sum_{i=1}^{N_f}\left[\bar{\scriptstyle D}^i_R Q_L^{\D_i}\right] S^{\D_i}_{R}\right) +
\nonumber\\
&+&
\lambda_S S_0 \left(\tilde{H}^{\dag} S^{\U_i}_L S^{\U_i}_R+ H^{\dag} S^{\D_i}_L S^{\D_i}_R\right)
+h.c.,
\label{LagMS}
\eea 
where contractions with color indices are understood and 
$S_0$ is a real singlet scalar field.
Here $q^i_L$, and ${\scriptstyle U}^i_R$, ${\scriptstyle D}^i_R$, 
indicate the SM fermion fields, and  $H$ is the SM Higgs  doublet, with
$\tilde{H}=i\sigma_2 H^{\star}$.
We do not report here the subdominant scalar terms needed to avoid the domain wall problem, see the discussion above.
We also do not report the expression for the interaction Lagrangian of the messenger scalar fields 
with the SM gauge bosons since the corresponding Lagrangian follows from the universal 
structure of gauge interactions. 
Furthermore, the messenger fields are also charged under $U(1)_F$ and carry the
same $U(1)_F$ charges as the correspondent dark fermions.

In principle, there is no reason why the masses
of the up and down-scalar messenger fields should be flavor independent. 
However, if one assumes that the only source of flavor breaking comes from the quantum charge sector, 
then imposing the flavor universality for the free Lagrangians in
the up- and down- scalar sector separately 
turns out to be a minimal and natural choice.
Unavoidably, the flavor breaking contained in the 
gauge sector is then communicated to the scalar sector at one loop level.
However, since this effect will be suppressed by $U(1)_F$ gauge coupling and loop effects, the flavor dependence in 
the messenger mass-sector should be considered as a small deviation from flavor universality. We will neglect this small effect in our analysis and 
assume, as a minimal choice, four flavor-universal free mass parameters 
$\tilde{m}_{U_L}$, $\tilde{m}_{U_R}$, 
$\tilde{m}_{D_L}$,  and $\tilde{m}_{D_R}$, corresponding to the mass
terms of the  
$S^{\U}_{L}$, $S^{\U}_{R}$, $S^{\D}_{L}$, and $S^{\D}_{R}$ fields, respectively.

As explained before, the following discrete symmetry $H\to -H$ and
 $S_0\to -S_0$ must be imposed to the whole Lagrangian in order to avoid 
tree level Yukawa couplings. 
However, in order to radiatively generate the SM Yukawa couplings
we have to require that the singlet scalar field $S_0$ acquires a VEV, 
namely $<S_0>=\mu$. There is no problem with the unwanted massless Goldstone 
boson in this case, since this is a discrete symmetry.

\begin{figure}[t]
\begin{center}
\includegraphics[width=0.4\textwidth]{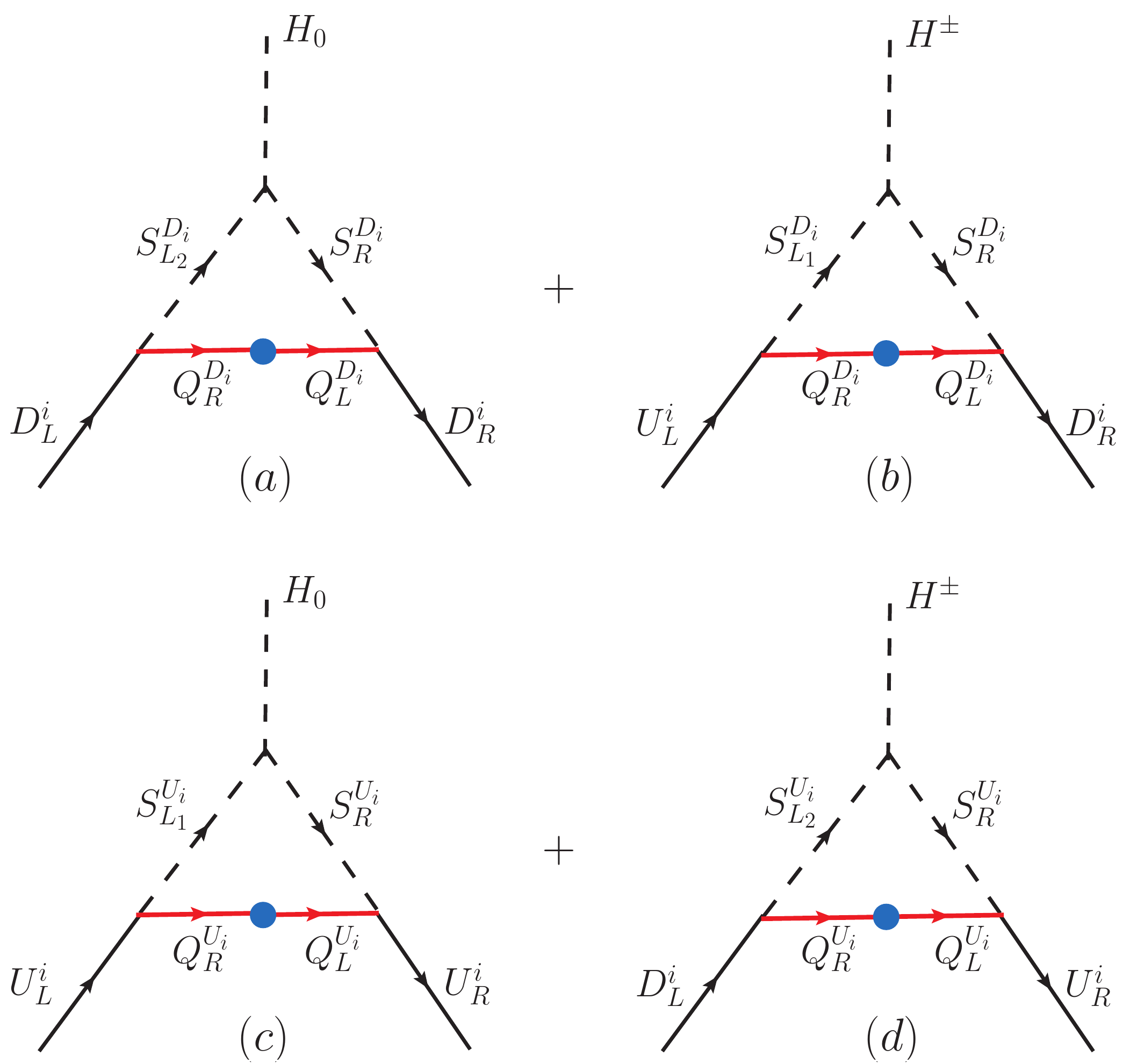}
\caption{One-loop contributions to the Higgs Yukawa couplings of down-quarks 
(a),(b) and up-quarks (c), (d). The internal dashed- and (red) continuous-lines 
stand for the scalar-messenger fields and dark-fermion fields respectively, 
while the dark (external) continuous lines indicate the quark fields. Underscore 
$L,R$ on the external quark fields stand for the corresponding chirality
projections. The external dashed lines correspond to the $SU(2)_L$ Higgs 
components $H_0$ and $H^{\pm}$.}
\label{fig2}
\end{center}
\end{figure}

In Fig.~\ref{fig2} we show the relevant Feynman diagrams which contribute to the SM Yukawa couplings at one loop order.
These diagrams are finite at one loop order, and in general at 
any order in perturbation theory, due to the structure of the renormalizable interaction in Eq.(\ref{LagMS}) and  the SSB of the discrete parity symmetry 
$H\to-H$ and $S_0\to -S_0$.

By computing the Feynman diagrams in Fig.~\ref{fig2}, the SM Yukawa couplings 
at zero transferred momenta can be
extracted by using the standard procedure as follows.
We match the results of the Feynman diagrams in Fig.~\ref{fig2}, where
the external  momenta are set to zero,  
with the corresponding effective Yukawa  operators evaluated at $q^2=0$.
In the calculation of the one loop diagrams we assume for simplicity
that the masses of the 
scalar fields running in the loop are flavor independent and their masses $\bar{m}$ are degenerate between the left and right scalars. 
Finally, by following the above procedure, we get
\bea
Y^{\U_i}&=& \frac{ \lambda_S \, g_L \, g_R \, \mu \, M_{Q^{\U_i}} }
{16 \pi^2\, \bar{m}^2}  C_0(x_i) \, ,
\label{Yukawa1}
\eea
and analogously for the $Y^{\D_i}$ sector, where $x_i=M_{Q^{\U_i}}^2/\bar{m}^2$
and $M_{Q^{\U_i}}=\Lambda\exp{\left(-\frac{2\pi}{3 \alpha q_{\U_i}^2}\right)}$, 
where $\alpha$ stands for the fine structure constant of $U(1)_F$ gauge 
interaction.
Here the function $C_0(x)$ is defined as
\bea
C_0(x)=\frac{1-x\left(1-\log{x}\right)}{(1-x)^2}\, ,
\label{C0}
\eea
where $C_0(1)=1/2$, while for small $x\ll 1$ it can be approximated as 
$C_0(x)\simeq 1+(1+\log{(x)})x +{\cal O}(x)$. In the opposite limit of 
large $x\gg 1$, one has $C_0(x)\sim 1/x$.
Therefore, from these results we can see that, as expected from the decoupling theorem, in the limit of $\Lambda \to \infty$ all the Yukawa couplings tend to zero.

Finally, after EWSB, the SM fermions get the same mass pattern as in Eq.(\ref{mgap2}), as in the example discussed in section II, namely
\bea
m_i &=& \Lambda_{\rm eff}\exp{\left(-\frac{2\pi}{3 \alpha q_{i}^2}\right)}
\eea
where now $q_i$  is the corresponding $U(1)_F$  charge of the 
corresponding dark fermion partner, 
and the $\Lambda_{\rm eff}$ is related to the 
Lee-Wick scale $\Lambda$ of $U(1)_F$ by
\bea
\Lambda_{\rm eff}&\sim &\left(\frac{v\mu \Lambda\, }{ \bar{m}^2}\right)
\frac{\lambda_S g_Lg_R C(x_i)}{16 \pi^2}
\, ,
\eea
with $v$ the Higgs vev,  $\bar{m}$ an average mass of the associated messengers fields and $g_{L,R}$
the corresponding messenger couplings to SM left-handed and right-handed 
fermions.
 Since there is no reason why the messenger scalar fields in the lepton and quark sectors should have the same mass and couplings, it is possible to choose
different masses and couplings $g_{L,R}$ for  the messengers fields in the lepton and quark sector, in order to set the appropriate scales 
$\Lambda_{\rm eff}$ for the lepton and quark sectors.

In the case in which the messenger sector is flavor independent
one can 
obtain interesting sum rules that connect the various Yukawa couplings
in 
the up- or down-quark sectors.
By means of Eqs.(\ref{mgap2}), (\ref{Yukawa1}) we get
\bea
Y^{\U_j}=Y^{\U_i} \, \exp{\left\{\frac{2\pi\left(q_{\U_j}^2-q_{\U_i}^2\right)}
{3\alpha\,q_{\U_i}^2q_{\U_j}^2}
\right\}}
\frac{C_0(x_j)}{C_0(x_i)}\, .
\label{Yukawasrel}
\eea
Analogous results hold for the down-sector Yukawa couplings $Y^{\D_j}$, 
with  $q_{\U_i}$ charges replaced by the corresponding $q_{\D_i}$ ones.
Clearly, if the 
messenger sector is flavor universal in both up- and down sector, the
above relations in Eq.(\ref{Yukawasrel}) can be generalized to mix the up- and 
down-sector  Yukawa couplings. 
As explained above, in order to avoid stable heavy charged particles in the spectrum, messenger masses should be always heavier than the corresponding
dark fermion ones. In the case of a large mass gap between the messenger and 
dark fermion sector, the last term multiplying the exponential 
in Eq.(\ref{Yukawasrel}) can be well approximated by $C_0(x_j)/C_0(x_i)\sim 1$.
In the following we will restrict our phenomenological analysis to this particular scenario.

The next issue to address  is the origin of flavour mixing.   In our framework the generated Yukawa couplings are proportional to the fermion masses in the dark sector.
There are two logical possibilities, either the observed flavour mixings are present already among the dark fermions or, alternatively, they are generated by the radiative
transfer mechanism. The first possibility requires dynamical breaking of the dark $U(1)_F$ symmetry since the 
charge conservation requires the mixing to be either zero (different $U(1)_F$
charges for different generations) or maximal (same $U(1)_F$ charges for different generations). As long as the dark photon acquires a small mass, much smaller than the 
generated fermion masses, the exponential dependence of masses on the $U(1)_F$ quantum numbers is not spoiled~\cite{Gabrielli:2007cp}. However, such a dynamical breaking
requires an additional mechanism and we do not consider it here. Instead, we assume that the small Cabibbo-Kobayashi-Maskawa (CKM) 
 type mixings are due to a mismatch between the dark sector
masses and the SM masses. Thus, they are generated by the scalars mediating the dark  fermion masses to the SM sector. To achieve that, we have to relax the
assumption of flavour universality of the messenger sector. However, due to the smallness of CKM mixing angles 
this is just a small mismatch effect originating from the flavour non-diagonal
messenger couplings and from the messenger mass non-universality. Thus the CKM matrix can always been accommodated in our mechanism.

Finally, we would like to comment about the phenomenological 
implications of the spontaneous ChSB in the dark sector. 
In the case of degenerate $U(1)_F$ charges, 
there is a global symmetry of the Lagrangian in the dark sector  
which corresponds to $U(N)_R\times U(N)_L$. After the spontaneous ChSB, induced by the higher derivative term in the $U(1)_F$ gauge sector, this symmetry breaks down to an exact $U(N)_V$ global symmetry. According to the Nambu-Goldstone 
theorem, there should then appear in the spectrum $N^2$ massless 
Nambu-Goldstone pseudoscalar bosons, 
that in this case would correspond to the
condensates of elementary dark fermions. Some of these composite
states would be also charged under the $U(1)_F$ gauge group.
On the other hand, if the  $U(1)_F$ charges are all non-degenerate, 
the $U(1)_F$ gauge interaction term will play the role of an explicit 
$SU(N)_L\times SU(N)_R$ breaking term. Then, according to general arguments, we
expect that of the $N^2$ Nambu-Goldstone particles of the degenerate case, 
only one will remain massless, 
while the other $N^2-1$ ones will acquire a mass term 
proportional to the splitting of the  $U(1)_F$ charges. 
Clearly, a rigorous analysis is mandatory in order confirm these naive 
expectations,  and this might be the subject for future investigations.

\section{Naturalness and vacuum stability bounds}

The radiative generation of the Yukawa couplings of light quarks has already been 
extensively considered in the literature in the context of supersymmetry~\cite{Banks:1987iu,ArkaniHamed:1995fq,ArkaniHamed:1996zw,ArkaniHamed:1996xm,Borzumati:1999sp}. 
In this framework the radiative generation of the top quark mass was considered to be
impossible because  the supersymmetry breaking scale was believed to be below 1~TeV, and generating a particle mass of 173~GeV at 
one loop seems impossible. As already discussed above, in our case we can choose the singlet VEV $\mu$ and the mass scales 
large enough to overcome the smallness of the loop factor.  Thus all SM Yukawa couplings can be generated with our mechanism.

However, large values of $\mu$, required to generate the top-quark Yukawa coupling,  can in principle spoil naturalness in the Higgs sector. This is due to the fact that the trilinear coupling of the Higgs and messenger sector can induce 
one loop contributions to the Higgs mass square $\delta m^2_{H}$, 
which is of order 
\bea
\delta m^2_{H}\sim \frac{\lambda_S^2\mu^2}{16\pi^2}\, .
\label{deltaH}
\eea
In this expression we have neglected the loop function since
we are just interested in a rough estimate of the contribution to 
the Higgs boson mass.
By using  Eqs.(\ref{deltaH}) 
and (\ref{Yukawa1}), and approximating the top Yukawa coupling by $Y^t\sim 1$,
the one loop radiative contribution
to the Higgs mass square $\delta m_H^2$ is given by
\bea
\delta m_H^2 &\sim& \frac{16 \pi^2 \bar{m}^2}{(g_Lg_R)^2 \, x_t C^2_0(x_t)},
\label{dmh1}
\eea
where  $x_t= M_{Q^t}^2/\bar{m}^2$. 
From these results we can see that in order to avoid large fine-tuning in the Higgs sector, large couplings of $g_L$ and $g_R$ are needed. 
Contrary to the radiative generation of Yukawa couplings in SUSY models, 
in our framework the messenger couplings to the Higgs boson are not constrained by any symmetry, and we can allow the $g_{L,R}$ couplings to be large. If
we assume that 
the mass of the dark fermion partner of the top-quark is of the same order as the messenger mass scale $\bar{m}$, namely $x_t\sim 1$, and assume 
$g_{L,R}\sim 1$, we get 
\bea
\delta m_H^2 &\sim & 4\times 10^{4}\left(\frac{\bar{m}}{{\rm TeV}}\right)^2 \, 
m_H^2 \, ,
\eea
for the Higgs mass $m_H=126$ GeV. This implies that for the
messenger mass scale of order $\bar{m}\sim 1$ TeV, a $10^{-4}$ 
fine-tuning  is required in the Higgs sector.

A potential solution to the fine tuning problem 
might be provided by extending the Lee-Wick ghosts to the SM fields, 
including the Higgs field, which is actually one of the main motivations for this proposal~\cite{Grinstein:2007mp,Grinstein:2008qq,Grinstein:2008bg,Espinosa:2011js}.
Another possibility is to consider the supersymmetric extension of our scenario,
that would necessarily require also the supersymmetric extension of the dark 
sector.

Now we derive the lower bounds on the dark fermion masses by using
vacuum stability bounds in the messenger scalar sector.
In order to simplify the analysis, we assume 
the messenger masses to be degenerate, that is 
$m_{S_L} \sim m_{S_R} \equiv \bar{m}$.
After electroweak symmetry breaking  the interaction term 
$\lambda_S\mu H S_L S_R$ generates a mixing term in the mass-square matrix of 
the $S_L$ and $S_R$ scalar fields, which is equal to
$\lambda_S\mu v S_L S_R$. If this mixing term is too large, 
one of the eigenvalues of the scalar mass-square matrix 
becomes negative and tachyons are generated, inducing vacuum instability.
Then, in order to avoid tachyons in the messenger sector, we must require that 
\bea
\lambda_S\mu v < \bar{m}^2,
\label{eq1}
\eea
where $v$ is the VEV of the Higgs field.

The SM fermion masses are generated as in the SM after the electroweak symmetry breaking.  We get from Eq.(\ref{Yukawa1})
\bea
\frac{m_i}{v} = \frac{L \lambda\mu M_{Q_i}}{\bar{m}^2}C_0(x_i),
\label{eq2}
\eea
where  $m_i$ is the SM fermion mass, $x_i=M_{Q_i}^2/\bar{m}^2$, and
for simplicity we absorbed in the constant $L$ all loop factors and
coupling constants, namely $L \sim g_L g_R/(16 \pi^2)$.
Now, from Eq.(\ref{eq2}) we get 
\bea
\lambda_S\mu= \frac{m_i \bar{m}^2}{v L M_{Q_i} C_0(x_i)}.
\label{eq3}
\eea
Substituting Eq.(\ref{eq3}) into Eq.(\ref{eq1}) we get
\bea
M_{Q_i} > \frac{m_i}{L C_0(x_i)}\, ,
\label{bounds}
\eea
that provides a lower bound on the dark fermion mass 
in terms of the corresponding SM fermion partner. Notice that, in the 
case of heavy messengers ($x_i\ll 1)$,  the lower bound depends 
only on the fermion masses and coupling constants.

Some comments about relation Eq.(\ref{bounds}) are in order.
In the case in which the dark fermion associated with the 
top-quark has  a mass of the same order as the messenger ones, namely 
$x_t\sim 1$, we get 
\bea
M_{Q_t} \gsim \left(\frac{55}{g_Lg_R}\right)\, {\rm TeV}\, .
\label{bound1}
\eea
In the case of large couplings $g_{L,R}\sim 1$, but still perturbative, 
the heaviest dark fermion should have a mass not smaller than 55 TeV to avoid
problems with the vacuum stability. On the other hand, for the lightest quarks, assuming their mass to be of order 10 MeV, we get 
\bea
M_{Q_u} \gsim \left(\frac{1.6}{g_Lg_R}\right)\, {\rm GeV}\, .
\label{bound2}
\eea
Clearly, if  the masses of the dark fermions are just a rescaling of the 
the SM fermion masses, as suggested by our scenario, the bound  Eq.(\ref{bound2}) automatically holds once the bound Eq.(\ref{bound1}) is satisfied.

These results show that the lightest dark fermions could be relatively light 
for strongly coupled messenger fields and can be produced at the LHC
in the decays of (heavy) messenger fields. However, in order for the messengers to be kinematically accessible at 
collider experiments, the bound Eq.(\ref{bound1}) should be relaxed.  It is possible that the messenger masses for the quarks and leptons are different.
For the lepton partners the equivalent bound is rescaled by the ratio of Yukawa couplings squared, allowing them to be kinematically reachable at colliders.
We will discuss  the phenomenological implications of this scenario at the LHC in the next section.

Finally, we would like to comment on the fact that this scenario can easily 
pass all the tests from electroweak precision observables and flavor physics. 
For instance, due to the fact that the messenger fields are charged under the 
$SU(1)_L\times U(1)_Y$ gauge group, they can contribute at one loop level to 
 the $\rho$ parameter. However, since the messenger masses may be as large as  50 TeV and also degenerate, we
expect them not to contribute significantly to the $\rho$ parameter and to 
the other electroweak precision observables. 
The same conclusions hold for the contribution to rare processes in  flavor physics induced at one loop. Since 
the messenger fields enter in flavor-changing neutral current (FCNC) loops, this will induce a tiny contribution to the relevant FCNC operators, being  suppressed by a typical scale which should be associated with the messenger masses. However, due to the fact that $g_{L,R}$ might be large,  an accurate analysis of these new contributions to the FCNC sector is needed in order to assess this issue more precisely.

\section{Phenomenology and direct tests }
The dark sector of our theory contains an unbroken $U(1)_F$ gauge group. Thus there must exist massless dark photons that may have 
cosmological implications if the dark matter of the Universe is charged under this gauge group~\cite{Ackerman:2008gi}. 
Recently there has been a revival of  interest to this possibility~\cite{An:2013yfc}.
The dark matter self interactions may solve problems of small scale structure 
formation that seem to deviate from the simple N-body simulation results. Spectacular signatures of this scenario include formation of 
dark discs of galaxies~\cite{Fan:2013tia} that can be observable. 
If the dark photon is exactly massless, there is no kinetic mixing with the electromagnetic photon -- there are two orthogonal states that 
must be identified accordingly.
In our scenario the natural candidate for dark matter is the lightest dark fermion that is charged under the $U(1)_F$ gauge group.
If, however, the dark matter is neutral under $U(1)_F$, the dark photons are very difficult to observe in laboratory experiments.

While probing the dark sector particles at colliders is a very challenging task, neutrino physics may offer unexpected possibilities.
Namely, some of the dark fermions, for example the ones corresponding to the lightest SM neutrinos, may be light enough to play the role of an additional sterile neutrino.
The existence of ${\cal O}(10)$~eV sterile neutrinos may be hinted by  the LSND~\cite{Aguilar:2001ty}  and Mini-BooNE~\cite{AguilarArevalo:2010wv} experiments. 
To mix the dark and the SM fields, the dark gauge symmetry must be broken. Thus the phenomenology of our scenario may also affect neutrino physics. 
However, it is not yet clear if this simple scheme, which would assume Dirac neutrinos, could explain the correct scales for the neutrino 
masses and mixing. A more close inspection of this model in the neutrino
sector is necessary and this is beyond the purpose of the present paper. 
Perhaps extended versions could be be necessary in the neutrino sector to make this model more realistic.

However, by far the most promising way to test our model is to search for
direct or indirect effect of this scenario at the LHC and in future colliders. 
As already stated, the messengers themselves must carry  SM quantum numbers similarly to the squarks and sleptons of supersymmetric theories. 
 If kinematically accessible, those new particles can be produced and discovered at the LHC. 
For example, the coloured messengers can be pair produced at the LHC by the
gluon fusion mechanism 
\bea
g\, g \to S \, S^{\dag}\, ,
\label{p1}
\eea
or by the quark fusion mechanism
\bea
q\, \bar{q} \to S\,  S^{\dag} \, ,
\label{p2}
\eea
where $S$ stands for a generic scalar messenger.
The latter process can be enhanced by the potentially large $g_{L,R}$ couplings.
For colourless messengers only the process \rfn{p2} can take place, mediated by the SM gauge bosons. 
This phenomenology would somewhat resemble the one of supersymmetric squarks and sleptons. However, 
there are many differences between those scenarios.
While in supersymmetric theories searches for squarks assume that they are produced in gluino cascade decays, our model does not contain coloured fermions and the scalars must be produced directly. This implies  smaller cross sections and lower mass reach than in supersymmetry, especially for colourless particles like the messengers of leptons.
Moreover, the masses of messenger fields are expected to be large, as follows from  the bound Eq.(\ref{bound1}) coming form the very large top Yukawa coupling. 
 Therefore, it is likely that the quark messenger cannot be produced on-shell at the LHC (unless the flavour universality assumption is relaxed that we do not consider in this work). 
However, the lepton messengers can be much lighter and accessible at the LHC. Once produced, the colourless scalars decay to SM leptons and to dark fermions.
The lightest dark fermion is stable, manifested at collider experiments by the signature of missing energy. 
Thus the experimental signature of our scenario is a pair of SM leptons and large missing energy.
The latter can be used to trigger on the events at the LHC. 
Thus the LHC searches for supersymmetry could also be used to test our flavour model.

Due to the direct coupling of the messenger sector with the Higgs boson, effects on the $H\to \gamma \gamma$ and gluon-decay amplitude $H\to g g$ can affect
the present measurements of the 126 GeV Higgs-like resonance observed at the LHC. We study how the radiative Higgs decay rates can be used to set direct bounds on the masses of particles in the messenger and dark fermion sector. 
Since the present measurements are in good agreement with SM predictions, one can use these results to set indirect lower bounds on the new particle spectra.
In particular, the messenger fields could contribute to the $H\to \gamma \gamma$
amplitude at one loop, where inside the loop  the 
$S_L^{\U_i,\D_{i}}$ and $S_R^{\U_i,\D_{i}}$ fields are circulating, together with their potential
counterparts in the leptonic sector. By dimensional analysis, we estimate that
this contribution is proportional to 
\bea
A(H\to \gamma \gamma) = \frac{\lambda_S\mu\alpha}{\bar{m}^2 4 \pi} L_F 
\hat{F}_{\mu\nu}\hat{F}^{\mu\nu},
\eea
where $\bar{m}$ is the average messenger mass, $L_F$ is the loop function, which is expected to be of order ${\cal O}(1)$, 
and $\hat{F}_{\mu\nu}$ is the Fourier transform of the EM field strength.
Now, if we extract the $\lambda_S\mu$ term 
from the requirement of generating the top Yukawa coupling at the right scale
by using Eq.(\ref{Yukawa1}), we get that the
amplitude for $H\gamma \gamma$ will be of order
\bea
A(H\to \gamma \gamma)\sim \frac{4\pi\alpha}{g_Lg_R M_{Q_t}\, C_0(x_t)}L_F
\hat{F}_{\mu\nu}\hat{F}^{\mu\nu}\, ,
\eea
while the corresponding SM contribution is proportional to 
\bea
A(H\to \gamma \gamma)_{SM}\sim \frac{\alpha g}{4\pi m_W}L_F^{SM}
\hat{F}_{\mu\nu}\hat{F}^{\mu\nu}\, ,
\eea
where $m_W$ is the W-boson mass, $g$ the weak coupling, and $L_F^{SM}$ the corresponding SM loop function, which is  a term of order ${\cal O}(1)$.
Since the vacuum stability bounds are restrictive,  see Eqs.(\ref{bound1},\ref{bound2}),   the messenger contribution to $H\to\gamma\gamma$ 
is expected to be suppressed with respect to the SM one. The same
conclusions hold for the new contribution to the Higgs 
production mechanism by gluon-gluon fusion.
Notice, that these estimates  are based on pure dimensional analysis, and the precise calculation of the bounds from the Higgs boson analysis at the 
LHC would require 
a dedicated study of these effects that goes beyond the purpose of the present paper.

\vspace{0.5cm}

\section{Conclusions}
We have proposed a new paradigm for the dynamical generation of exponentially spread SM Yukawa couplings.
The new idea we advertise  is that exponentially spread fermion masses are generated non-perturbatively in the dark sector. The resulting chiral and flavour symmetry breaking is 
transferred to the SM via the messenger fields presented in Table~1. The important ingredient for our mechanism to work is the existence of Lee-Wick negative norm ghosts
in the dark sector, allowing the NJL type mechanism to be operative in the weak coupling regime of the theory and producing an exponential mass spectrum. 
The interaction that generates the non-perturbative effect is the unbroken dark  $U(1)_F$ gauge interaction. 
Since the Abelian group can have different integer or fractional charges for different generations,  
flavour symmetries are broken exponentially by the $U(1)_F$ charges. As a result, our mechanism offers a natural explanation to the observed 
SM fermion mass spectrum. If the light neutrinos will turn out to be Dirac particles, explaining the extreme smallness of their Yukawa couplings 
becomes natural in our framework.

We have presented an explicit model of flavour achieving those tasks. It contains a scalar messenger sector consisting of particles with 
the SM quark and lepton quantum numbers, thus resembling the supersymmetric squark and slepton sector. 
We have shown that, due to the large top Yukawa coupling, quark messengers must likely be very heavy. However, the lepton messengers can be orders of magnitude lighter.
If kinematically accessible, those particles can be discovered at the LHC, offering direct tests of our model.

\vspace{0.5cm}

\mysection{Acknowledgments}
We thank C. Spethmann for suggestions on the manuscript. 
E.G. would like to thank the PH-TH division of CERN for its kind 
hospitality during the preparation of this work.
This work was supported by the ESF grants  MTT59, MTT60,
by the recurrent financing project SF0690030s09, and by the European Union through the European Regional Development Fund.

\end{document}